\documentclass[aps,pra,twocolumn,showpacs,longbibliography]{revtex4-2}
\usepackage[plainpages=false,pdfpagelabels,colorlinks=true,linkcolor=red,urlcolor=blue,citecolor=blue,pdftitle={Title},pdfauthor={},pdfdisplaydoctitle=true,pdfduplex=DuplexFlipLongEdge]{hyperref}
\usepackage{amsmath, amsthm, amssymb}
\usepackage{array}
\usepackage{bbold} 
\usepackage{bm}        
\usepackage[dvipsnames,usenames]{color}
\usepackage{dcolumn}
\usepackage{float}
\usepackage{graphicx}  
\usepackage{lipsum}
\usepackage{mathrsfs}
\usepackage{mathtools,amssymb,graphicx,units,bm}
\usepackage{physics}%
\usepackage{soul} 
\usepackage{ulem} 

\graphicspath{{figures/}}

\emergencystretch=\maxdimen
\def\ave#1{\langle #1\rangle}

\newcommand{\ellv}{\ensuremath{\boldsymbol\ell}}
\newcommand{\iv}{\mathbf{i}}
\newcommand{\jv}{\mathbf{j}}
\newcommand{\kv}{\mathbf{k}} 

\newcommand{\phdag}{{\phantom{\dagger}}}
\newcommand{\qv}{\mathbf{q}}

\begin{document}

\normalem

\title{Effects of next-nearest neighbor hopping on the pairing and critical temperatures of the attractive Hubbard model on a square lattice
}
\author{Rodrigo A. Fontenele}
\author{Natanael C. Costa}
\author{Thereza Paiva}
\author{Raimundo R. dos Santos}
\affiliation{Instituto de F\'isica, Universidade Federal do Rio de
Janeiro Cx.P. 68.528, 21941-972 Rio de Janeiro RJ, Brazil}


\begin{abstract}
    The attractive Hubbard model plays a paradigmatic role in the study of superconductivity (superfluidity) and has become directly realizable in ultracold atom experiments on optical lattices. However, the critical temperatures, $T_{c}$'s, remain lower than the lowest temperatures currently achievable in experiments. Here, we explore a possible route to enhance $T_c$ by introducing an additional next-nearest-neighbor (NNN) hopping, $t^\prime$, in a two-dimensional square lattice. We perform sign-problem-free determinant quantum Monte Carlo simulations to compute response functions such as pairing correlation functions, superfluid density, and uniform spin susceptibility. Our results show that a judicious choice of $t^\prime$ can increase $Tc$ by up to $50\%$ compared to the case with only nearest-neighbor hopping. In contrast, the preformed pairs temperature scale, named pairing temperature, $T_p$, decreases with increasing $|t^{\prime}/t|$, which should represent a reduction of the pseudogap region, favoring a more BCS-like behavior at intermediate coupling. We further analyze the interacting density of states to characterize the transition from a pseudogap regime to a fully gapped superconducting state. These findings suggest that NNN hopping could be a viable route to increase $T_c$ to values closer to experimentally accessible temperature scales. 
\end{abstract}
\maketitle

\section{Introduction}

Recent progress in optical lattice experiments (OLE's) with ultracold fermionic atoms has been instrumental in unveiling properties of 
quantum matter models\,\cite{Jaksch05,Bloch08,Esslinger10,Koepsell2021,Greiner-AF}. 
Amongst these, the Hubbard model is the simplest one capturing the interplay between itinerancy and on-site interactions, $U$. 
In the context of materials, repulsive interactions, $U>0$, between fermions have their origin in screened Coulomb potentials \cite{Hubbard63}. 
By contrast, local fermion pairs may be formed in narrow-band systems due to a local attractive short-ranged effective interaction \cite{Micnas90}, $U<~0$. 
The latter case is usually referred to as the attractive Hubbard model (AHM), and has been useful in studying some unconventional aspects of superconductivity, such as pseudogaps below a certain temperature scale, $T_p$, and the crossover between the Bardeen-Cooper-Schrieffer (BCS) and Bose-Einstein condensation (BEC) limits, as the pairing interaction increases \cite{Randeria92,Bucher93,Randeria95a,Randeria14,Fontenele22}.  
In OLE's both the magnitude and sign of the interactions are fine-tuned by Feshbach resonances driven by an external magnetic field \cite{chin10}.
In addition, the advent of the quantum gas microscope for bosons \cite{Bakr09} and later for fermions \cite{quantum-gas-f-1,quantum-gas-f-2,quantum-gas-f-3,quantum-gas-f-4,quantum-gas-f-5} paved the way to visualize the atomic distribution on the lattice and to extract quantitative data, such as correlation functions \cite{Parsons2016,Boll2016,Cheuk2016,Mitra18,Hartke23}. 

A recent experiment for the repulsive Hubbard Model in OLE's has reached unprecedented low temperatures at the order of $T \sim 0.05 t/k_B$ \cite{criogenic}, inaugurating a new era of cryogenic temperatures within the optical lattices framework; 
$t$ is the hopping integral between nearest-neighbor sites, and $k_B$ is the Boltzmann constant, which we set to unity from now on. Recent Quantum Monte Carlo (QMC) simulations for the attractive case\,\cite{Fontenele22} have mapped the phase diagram for the square lattice, showing a maximum $T_c\approx 0.15 t$  around $\langle n\rangle \approx 0.87$  \cite{Fontenele22}. Although this $T_c$ lies above the cryogenic OLE's, studies of the AHM in this range have not yet been reported, being limited to $T \sim 0.45 t = 3 T_c$ \cite{Mitra18} and $T \sim 0.36 t = 2.4\, T_c$ \cite{Hartke23}. 
Notwithstanding this  steady experimental decrease in temperatures, the theoretical search for higher $T_c$'s is of interest.

Inspired mainly by the BCS prediction for the critical temperature for the AHM at weak coupling \cite{Micnas90},
\begin{equation}
    T_{c}^\text{BCS}= \widetilde{W} \exp\left[-1/D(0)|U|\right],
    \label{eq:Tc_BCS}
\end{equation}
where $\widetilde{W}$ is proportional to the bandwidth, $W$, 
and $D(0)$ is the density of states (DOS) at the Fermi level, we may seek mechanisms to increase $D(0)$, while keeping fixed both the fermionic density and $U$.
One such mechanism was explored in the determinant quantum Monte Carlo (DQMC) simulations \cite{Fontenele23, Fontenele24, Scalettar25}. For instance, Ref.\,\cite{Fontenele24} has shown that tuning the interlayer hopping in a bilayer may lead to $T_c$ enhancements by 50 to 70\% with respect to the monolayer; similar enhancements have also been obtained in simple cubic lattices, by tuning the hopping along, say the $z$ direction~\cite{Fontenele24}.
At this point it is worth mentioning that Eq.\,\eqref{eq:Tc_BCS} cannot be taken as face value for accuracy: it merely tracks the DQMC $T_c$'s as there is an unknown multiplication factor in $\widetilde{W}$ in Eq.\,\eqref{eq:Tc_BCS}; see the discussion in Ref.\,\onlinecite{Fontenele24}. 

\begin{figure}[t]
  \centering
  \includegraphics[scale=0.5]{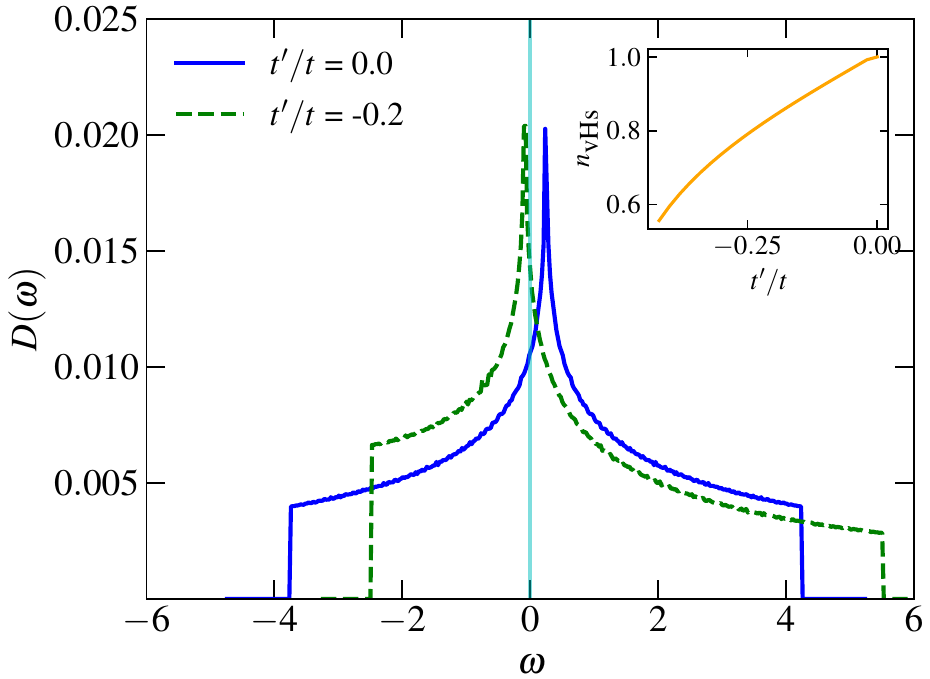}
   \caption{Non-interacting density of states as a for  $t^{\prime} = 0$ and $t^{\prime} = -0.2t$.
   The vertical lines locate the corresponding Fermi energies for $\ave{n}=0.87$.
   The inset shows the electronic density at which the van Hove singularity occurs, $n_{\mbox{vHS}}$, as a function of  $t^{\prime} /t$.}
 \label{fig:DOS}
\end{figure}

Here we examine yet another possibility to increase $T_c$, while keeping the system two-dimensional: we allow for both nearest and next-nearest neighbor (NNN) hopping, $t^\prime$, so that fermions may also move along the diagonals of a square lattice. 
This provides an additional channel for Cooper pairs to move around, avoiding pair-breaking scattering events by lattice sites. 
Indeed, Fig.\,\ref{fig:DOS} shows the non-interacting density of states for $\ave{n}=0.87$ (the density for which $T_c$ in the AHM is maximum over a range of values of $U$), for both $t'=0$ and $t'=-0.2t$; the corresponding Fermi levels sit at $\omega \equiv \varepsilon(\kv)-\mu=0$, with the dispersion $\varepsilon(\kv)$ given by Eq.\,\eqref{eq:ek}, and are indicated by the vertical lines. 
We see that for this filling $D(0)$ is larger for $t'=-0.2t$, which, according to Eq.\,\eqref{eq:Tc_BCS}, suggests a $T_c$ larger than for $t'=0$.
This picture is supported by perturbative approaches~\cite{hirsch86} and by early QMC simulations for limited sets of parameters \cite{dosSantos92, Fontenele23}, both showing an enhancement of superconducting pairing correlations for a finite $t^{\prime}$; we note, however, that these studies provide no indications about how high $T_c$ can reach.

\begin{figure}[t]
  \centering
  \includegraphics[scale=0.45]{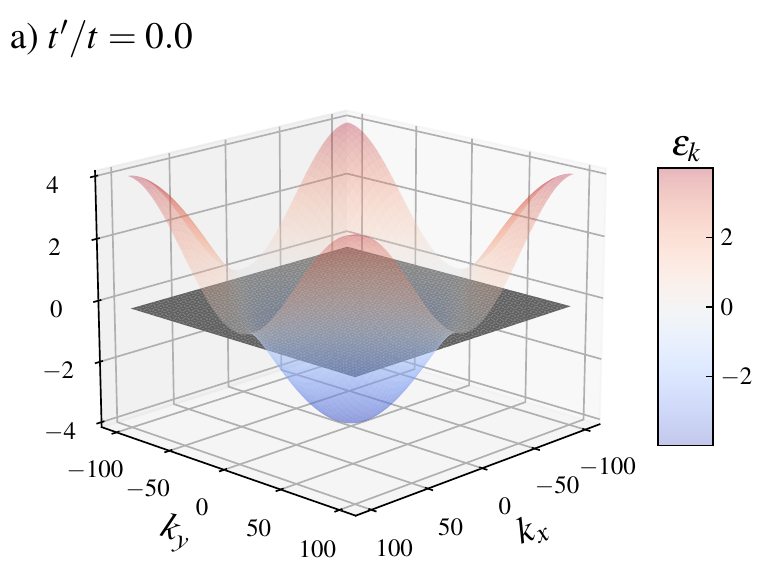}
  \includegraphics[scale=0.45]{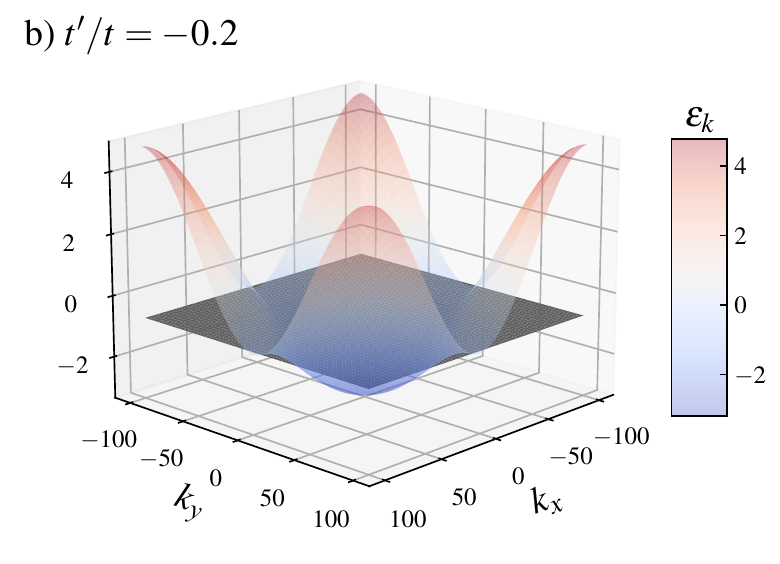}
   \caption{Band dispersion, Eq.\,\eqref{eq:ek} for (a) $t'/t = 0.0$ and (b) $t'/t = -0.2$. The gray planes locate the Fermi energy for $\langle n \rangle = 0.87$.}
 \label{fig:Ek}
\end{figure}

One should keep in mind that when $t'=0$ in two dimensions, a finite $T_c$ only occurs in the doped case, since the degeneracy between charge ordering and superconductivity at half filling causes a suppression of $T_c$ to zero, by virtue of the Mermin-Wagner theorem \cite{Micnas90}.
In what follows, we also address the possibility of $t'\neq 0$ inducing a finite $T_c$ even at half filling.  
The aforementioned experiments on the repulsive case at record-low temperatures \cite{criogenic} may open a viable route to explore superconductivity in the attractive counterpart, where a finite $t'$ is expected to enhance $T_c$ to experimentally accessible regimes.Further, with second neighbor hopping, the non-interacting band can become flatter than for $t'=0$.
Indeed, Fig.\,\ref{fig:Ek} shows the single-particle band, 
\begin{align}
   \varepsilon(\mathbf{k}) = -2t\left[\cos{k_{x}} + \cos{k_{y}} \right] - 4t^{\prime}\cos{k_{x}}\cos{k_{y}},
    \label{eq:ek}
\end{align}
and the flattening for  $t'=-0.2 t$ is significant.
Flat bands tend to enhance electronic correlations in the presence of interactions, thus giving rise to a more robust superconducting state \cite{Derzhko15, Tovmasyan16, Shaginyan19, Huang19}. The interplay between flat band and the emergence of superconductivity has been experimentally explored in materials such as magic angle twisted bilayer graphene, where unconventional superconductivity might be observed, through tunneling and Andreev
reflection spectroscopy, as a consequence of the large density of states at its flat bands \cite{Cao18, Lu19, Oh21, Carlon25}.

Still from an experimental point of view, we note that a triangular optical lattice has been set up experimentally by adding an extra pair of counter-propagating laser beams; see Ref.\,\cite{schauss21}. Given that the triangular lattice may be thought of as a square lattice with hopping along one of the diagonals, a possible route to generate NNN hopping on a square lattice would be to add yet another pair of counter-propagating laser beams, this time along the other diagonal direction. 
In view of these possibilities, a systematic study of the effects of NNN hopping on the critical temperature of the AHM is of interest. 
With this in mind, here we use DQMC simulations to map out how much $T_c$ can be enhanced by second neighbor hopping for different values of the attractive interaction and band filling. 
In addition, we also explore the behavior of the pairing temperature with $t'$. 


The layout of the paper is as follows. 
In Section \ref{sec:HQMC}, we discuss the model and highlight the main aspects of DQMC, including the quantities used to probe the physical properties of the system. 
In Section \ref{sec:result} we present our main results for the critical temperature and for the pairing temperature as functions of the NNN hopping amplitude.
Section \ref{sec:conc} summarizes our findings.

\section{Model and Methodology}
\label{sec:HQMC}

\begin{figure}[t]
  \centering
  \includegraphics[scale=0.45]{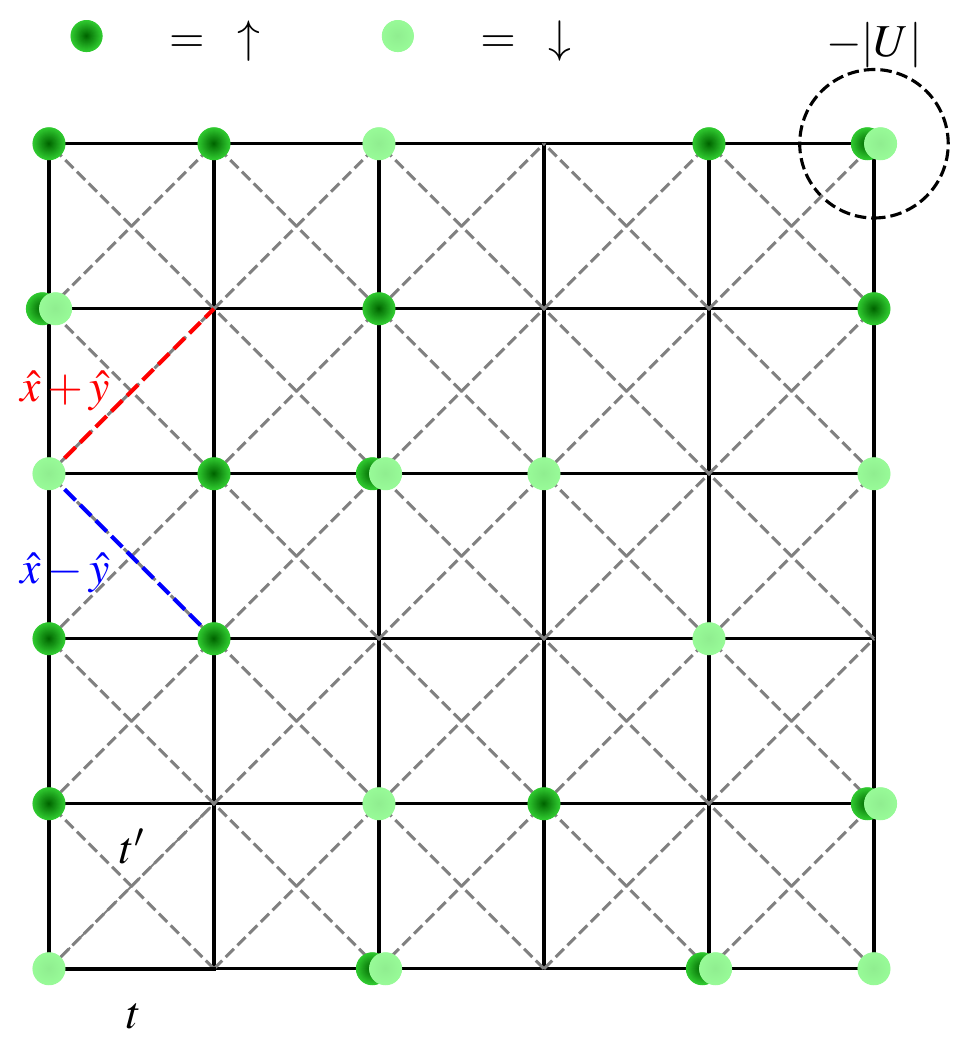}
   \caption{Typical portion of a square lattice illustrating the model parameters. 
   Two fermions with opposite spins on the same site contribute with $-|U|$ to the total energy.
   Fermions hopping between nearest-neighboring sites (horizontally, $\hat{\mathbf{x}}$, or vertically, $\hat{\mathbf{y}}$) contribute with $-t$ (full lines) to the total energy.
   Fermions hopping along the diagonals, $\hat{\mathbf{x}}\pm\hat{\mathbf{y}}$, contribute with $-t'$ (dashed lines) to the total energy.
   }
 \label{fig:squaLattice}
\end{figure}

The AHM Hamiltonian can be written, in standard second-quantized form, as
\begin{align}
\mathcal{H} &= -t\sum_{\langle \mathbf{i}, \mathbf{j}\rangle,\sigma}\left(c_{\mathbf{i}\sigma}^{\dagger}c_{\mathbf{j}\sigma}^\phdag + \mbox{h.c.}\right) - t^{\prime}\sum_{\langle\langle \mathbf{i}, \mathbf{j}\rangle\rangle,\sigma}\left(c_{\mathbf{i}\sigma}^{\dagger}c_{\mathbf{j}\sigma}^\phdag + \mbox{h.c.}\right) \nonumber \\
&~~ -\mu\sum_{\mathbf{i}\sigma}n_{\mathbf{i}\sigma} - |U|\sum_{\mathbf{i}}\left(n_{\mathbf{i}\uparrow}-1/2 \right)\left(n_{\mathbf{i}\downarrow}-1/2 \right),
\label{eq:3D_Hamilt}
\end{align}
where $\langle \mathbf{i}, \mathbf{j}\rangle$ and $\ave{\ave{\iv,\jv}}$ respectively restrict hoppings between  nearest- and next-nearest neighboring sites on a square lattice, with  amplitudes $t$ and $t'$ (see Fig.\,\ref{fig:squaLattice}); `h.c.'\ stands for Hermitian conjugate of the previous expression. 
The band filling is controlled by the chemical potential, $\mu$, and $|U|$ is the magnitude of the on-site attractive interaction between fermions of opposite spin. 
Throughout the paper we  take a unit lattice spacing, and the nearest neighbor hopping integral, $t$, sets the energy scale.

We investigate the finite temperature properties of the model through Determinant Quantum Monte Carlo (DQMC) simulations \cite{Blankenbecler81,hirsch83,hirsch85,white89,dosSantos03b}.
This is an unbiased numerical approach based on an auxiliary-field decomposition of the interaction, which is mapped onto a quadratic form of free fermions coupled to bosonic degrees of freedom, $\mathcal{S}(i,\tau) = \pm 1$, in both spatial and imaginary time coordinates.
The non-commuting parts of the Hamiltonian are separated through a Trotter-Suzuki decomposition, i.e.
\begin{align}
    	\mathcal{Z} &= \mathrm{Tr}\,e^{-\beta\widehat{\mathcal{H}}}
	= \mathrm{Tr}\,[(e^{-\Delta\tau(\widehat{\mathcal{H}}_{0} + \widehat{\mathcal{H}}_{\rm
U})})^{M}]
\thickapprox \nonumber \\
& \thickapprox \mathrm{Tr}\,[e^{-\Delta\tau\widehat{\mathcal{H}}_{0}}e^{-\Delta\tau\widehat{\mathcal{H}}_{\rm
U}}e^{-\Delta\tau\widehat{\mathcal{H}}_{0}}e^{-\Delta\tau\widehat{\mathcal{H}}_{\rm
U}}\cdots], 
\end{align}
where $\widehat{\mathcal{H}}_{0}$ contains the terms quadratic in fermion creation and destruction operators,  and $\widehat{\mathcal{H}}_{\rm U}$ contains the quartic ones.
We take $\beta = M \Delta\tau$ , with $\Delta\tau$ being the grid of the imaginary-time coordinate axis. This decomposition leads to an error proportional to $(\Delta\tau)^{2}$, which can be systematically reduced as $\Delta\tau\to 0$. Throughout this work, we choose $\Delta\tau \leq 0.1$ (depending on the temperature), which is small enough to lead to systematic errors smaller than the statistical ones (from the Monte Carlo sampling). 
Finally, we stress that for the AHM the discrete Hubbard-Stratonovich transformation dealing with the quartic terms in $\widehat{\mathcal{H}}_{\rm U}$ leads to sign-free simulations\,\cite{hirsch83,hirsch85,white89,dosSantos03b}.

In order to probe the emergence of superconductivity, we 
calculate the $s$-wave pair correlation functions,
\begin{equation}
C_{\mathbf{i}\mathbf{j}}^\Delta \equiv 
	\frac{1}{2}\langle b_{\mathbf{i}}^{\dagger} b_{\mathbf{j}}^{\phantom{\dagger}} 
	+ \text{H.c.}\rangle,
\label{eq:Cbibj}
\end{equation}
with $b_{\mathbf{i}}^{\dagger} \equiv c_{\mathbf{i}\uparrow}^{\dagger} c_{\mathbf{i}\downarrow}^{\dagger}$ ($b_{\mathbf{i}}^{\phantom{\dagger}} \equiv c_{\mathbf{i}\downarrow}^{\phantom{\dagger}}c_{\mathbf{i}\uparrow}^{\phantom{\dagger}}$) corresponding to  creation (annihilation) of a pair of electrons at a given site $\mathbf{i}$.
Then, the Fourier transform of $C_{\mathbf{i}\mathbf{j}}^\Delta$ at $\mathbf{q}=0$ defines the \textit{s}-wave pair structure factor,
\begin{equation}
	P_{s}(\mathbf{q}=0)  = \frac{1}{N}\sum_{\mathbf{i}, \mathbf{j}}  C_{\mathbf{i}\mathbf{j}}^\Delta
\label{eq:Ps-def}
\end{equation}
with $N=L\times L$ being the number of sites of the lattice, where $L$ is the linear lattice size;  periodic boundary conditions (PBC) are imposed.

We obtain estimates for $T_c$ through the superfluid density, $\rho_s$.
This quantity can be expressed in terms of the current-current correlation functions as \cite{Scalapino92,Scalapino93},
\begin{equation}
\rho_s=\frac{D_s}{4\pi e^2} =
\frac{1}{4} [\Lambda^L - \Lambda^T],
\label{Ds}
\end{equation}
where $D_s$ is the superfluid weight, and
\begin{equation}
\Lambda^L \equiv \lim_{q_x \to 0} \Lambda _{xx} (q_x, q_y=0,\omega_n=0),
\end{equation}
and
\begin{equation}
\Lambda^T \equiv \lim_{q_y \to 0} \Lambda _{xx} (q_x=0, q_y,\omega_n=0),
\end{equation}
are, respectively, the limiting longitudinal and transverse responses, 
with
\begin{equation}
	\Lambda_{xx}(\qv, \omega_n)=
 		\sum_{\ellv} \int_0^\beta d\tau\, 
		e^{i\qv \cdot \ellv} e^{i \omega_n \tau} \Lambda_{xx}(\ellv,\tau),
\label{lambdaq}
\end{equation}
where $\omega_n=2 n \pi T$ is the Matsubara frequency, and 
\begin{equation}
\label{lambda}
\Lambda_{xx}(\ellv, \tau)= \langle j_x(\ellv, \tau) j_x (0,0) \rangle ~.
\end{equation}
Where $j_x(\ellv,\tau)$ is the $x$-component of the current density operator, $\mathbf{J}$, which can be defined as,
\begin{align}
    \mathbf{J}(\ellv) &= \frac{t}{2}\sum_{\overset{\mathbf{\hat{\delta}} \in \left\{ \mathbf{\hat{\delta}}\right\}}{\sigma}}\hat{\bf{\delta}}\left(i c_{\ellv + \hat{\bf{\delta}}, \sigma}^{\dagger}c_{\ellv, \sigma}^{\phdag} + \text{H.c.} \right) \nonumber \\
    &~ + \frac{t^{\prime}}{2}\sum_{\overset{\mathbf{\hat{\delta^{\prime}}} \in \left\{ \mathbf{\hat{\delta}^{\prime}}\right\}}{\sigma}}\hat{\bf{\delta}^{\prime}}\left(i c_{\ellv + \hat{\bf{\delta}^{\prime}}, \sigma}^{\dagger}c_{\ellv, \sigma}^\phdag + \text{H.c.} \right)~,
\end{align}
with $\left\{ \mathbf{\hat{\delta}} \right\}$ and $\left\{\mathbf{\hat{\delta^{\prime}}}\right\}$ including all unitary displacements for nearest and next-nearest neighbor sites, namely $\left\{ \mathbf{\hat{\delta}} \right\} = \left\{\pm \hat{x},~ \pm \hat{y} \right\}$ and $\left\{ \mathbf{\hat{\delta}^{\prime}} \right\} = \left\{\pm \hat{x}\pm \hat{y},~ -\hat{x}+\hat{y},~\hat{x}-\hat{y}\right\}$; see Fig.\,\ref{fig:squaLattice}. 
Thus, we can write
\begin{align}
    j_x(\ellv,\tau)= e^{ {\cal H} \tau} 
		\left[ \mathbf{J}(\ellv)\cdot\hat{x}
		 \right] e ^{-{\cal H} \tau};
\end{align}
see Ref.\,\cite{Scalapino92} for details.

Gauge invariance requires that the longitudinal part of $\Lambda$ satisfies the equality \cite{Scalapino92},
\begin{align}
    \Lambda^{L} \equiv \lim_{q_x \to 0} \Lambda _{xx} (q_x, q_y=0,\omega_n=0) = -K_{x}
    \label{eq:defKx}
\end{align}
where $K_{x}$ is the kinetic energy contribution relative to the $x$ direction, 
\begin{align}
    K_{x} &= \biggl< -t\sum_{\sigma}\left( c_{\ellv + \hat{\mathbf{x}},\sigma}^{\dagger}c_{\ellv,\sigma} + c_{\ellv,\sigma}^{\dagger}c_{\ellv + \hat{\mathbf{x}},\sigma} \right) \nonumber \\
    &~~ - t^{\prime}\sum_{\sigma}\left( c_{\ellv^{\prime} + \hat{\mathbf{x}}^{\prime}+\hat{\mathbf{y}}^{\prime},\sigma}^{\dagger}c_{\ellv^{\prime},\sigma} + c_{\ellv^{\prime},\sigma}^{\dagger}c_{\ellv^{\prime} + \hat{\mathbf{x}}^{\prime} +\hat{\mathbf{y}}^{\prime},\sigma} \right) \nonumber \\ 
    &~~ - t^{\prime}\sum_{\sigma}\left( c_{\ellv^{\prime} + \hat{\mathbf{x}}^{\prime} - \hat{\mathbf{y}}^{\prime},\sigma}^{\dagger}c_{\ellv^{\prime},\sigma} + c_{\ellv^{\prime},\sigma}^{\dagger}c_{\ellv^{\prime} + \hat{\mathbf{x}}^{\prime} -\hat{\mathbf{y}}^{\prime},\sigma} \right) \biggr>~,
\end{align}
in which we take into account both NN and NNN contributions.

At a KT transition, the following universal-jump relation involving the helicity modulus holds \cite{Nelson77}:
\begin{equation}\label{eq:helicity_mod}
T_c = \frac {\pi} {2} \rho_s^-,
\end{equation}
where $\rho_s^-$ is the value of the helicity modulus just below the critical temperature. We therefore calculate both $\Lambda^L$ and $\Lambda^T$ by DQMC simulations to obtain $\rho_s$ through Eq.\,\eqref{Ds}. 
$T_c$ is then determined by plotting $\rho_s(T)$, and looking for the intercept with $2T/\pi$ \cite{Denteneer91,Denteneer93,Denteneer94,Paiva04}; see Fig.\,\ref{fig:phos} below.

In addition, charge ordering is probed with the aid of the CDW structure factor,
\begin{equation}
    P_{\text{CDW}}(\textbf{q}) = \frac{1}{N} \sum_{\mathbf{i},\mathbf{j}} e^{i (\mathbf{i} - \mathbf{j}) \cdot \textbf{q}} \langle n_{\mathbf{i}} n_{\mathbf{j}} \rangle,
    \label{Eq:Ps_CDW}
\end{equation}
with the staggered CDW order corresponding to $\textbf{q} = (\pi, \pi)$.

We also discuss the behavior of a pairing temperature scale, $T_p$, which may be signaled by a gap opening for spin excitations, hence as a downturn in the uniform magnetic susceptibility $\chi_{s}$ as the temperature is lowered \cite{Micnas90,Randeria92,dosSantos94,Wlazlowski13,Tajima14,Fontenele22}.
The uniform susceptibility is obtained through the fluctuation-dissipation theorem, where $\chi_s = \beta \langle S^2 \rangle$, with $\langle S^2 \rangle$ being the uniform spin structure factor,
\begin{equation}
\langle S^2 \rangle =\frac{1}{3N}\sum_{\mathbf{i}\mathbf{j}} \ave{\mathbf{S}_{\mathbf{i}}\cdot\mathbf{S}_{\mathbf{j}}},
\label{eq:magsusc}
\end{equation}
where $\mathbf{S}_{\mathbf{i}}\equiv (1/2)\mathbf{m}_{\mathbf{i}}$, with the components of the magnetization operator being
\begin{subequations}
\label{eq:mags}
\begin{eqnarray}
m_{\mathbf{i}}^x&\equiv&
c_{\mathbf{i}\uparrow}^\dagger c_{\mathbf{i}\downarrow}^{\phantom{\dagger}}+ c_{\mathbf{i}\downarrow}^\dagger c_{\mathbf{i}\uparrow}^{\phantom{\dagger}},
\label{eq:mx}\\
m_{\mathbf{i}}^y&\equiv&
-i\left(c_{\mathbf{i}\uparrow}^\dagger c_{\mathbf{i}\downarrow}^{\phantom{\dagger}}- c_{\mathbf{i}\downarrow}^\dagger c_{\mathbf{i}\uparrow}^{\phantom{\dagger}}\right),
\label{eq:my}\\
m_{\mathbf{i}}^z&\equiv&n_{\mathbf{i}\uparrow}-n_{\mathbf{i}\downarrow}.
\label{eq:mz}
\end{eqnarray}
\end{subequations}

Our simulations were carried out on $L\times L$ lattices, with $L\leq 16$. 
Typically, our data have been obtained after $5-10\times 10^{3}$ warming-up steps, followed by $2-6\times 10^{5}$ sweeps for measurements, depending on the temperature, interaction strength, and electronic density.

\section{Results}
\label{sec:result}

\subsection{The case of a half-filled band}
\label{subsec:halffilling}

To analyze the half filling case, it is important to recall some symmetries of the attractive Hubbard model. These become evident when expressed in terms of the pseudo-spin operators \cite{Anderson1958,Auerbach2012},
\begin{align}
\eta^{z} & = \frac{1}{2}\sum_{\mathbf{i}} \left( n_{\mathbf{i}\uparrow} + n_{\mathbf{i}\downarrow} - 1 \right) \\
\eta^{+} & = \left( \eta^{-} \right)^{\dagger} = \sum_{\mathbf{i}} \epsilon_{\mathbf{i}}^{\phantom{\dagger}} c^{\dagger}_{\mathbf{i}\uparrow} c^{\dagger}_{\mathbf{i}\downarrow}\,
\end{align}
where $\epsilon_{\mathbf{i}} = \pm 1$ depending on whether site $\mathbf{i}$ lies on sublattice A or B. In this formulation, $\eta^{z}$ encodes charge fluctuations, while $\eta^{\pm}$ generates pairing; together, they define the one-component CDW and the two-component SS order parameters, respectively.
\begin{figure}[t]
\centering
\includegraphics[scale=0.44]{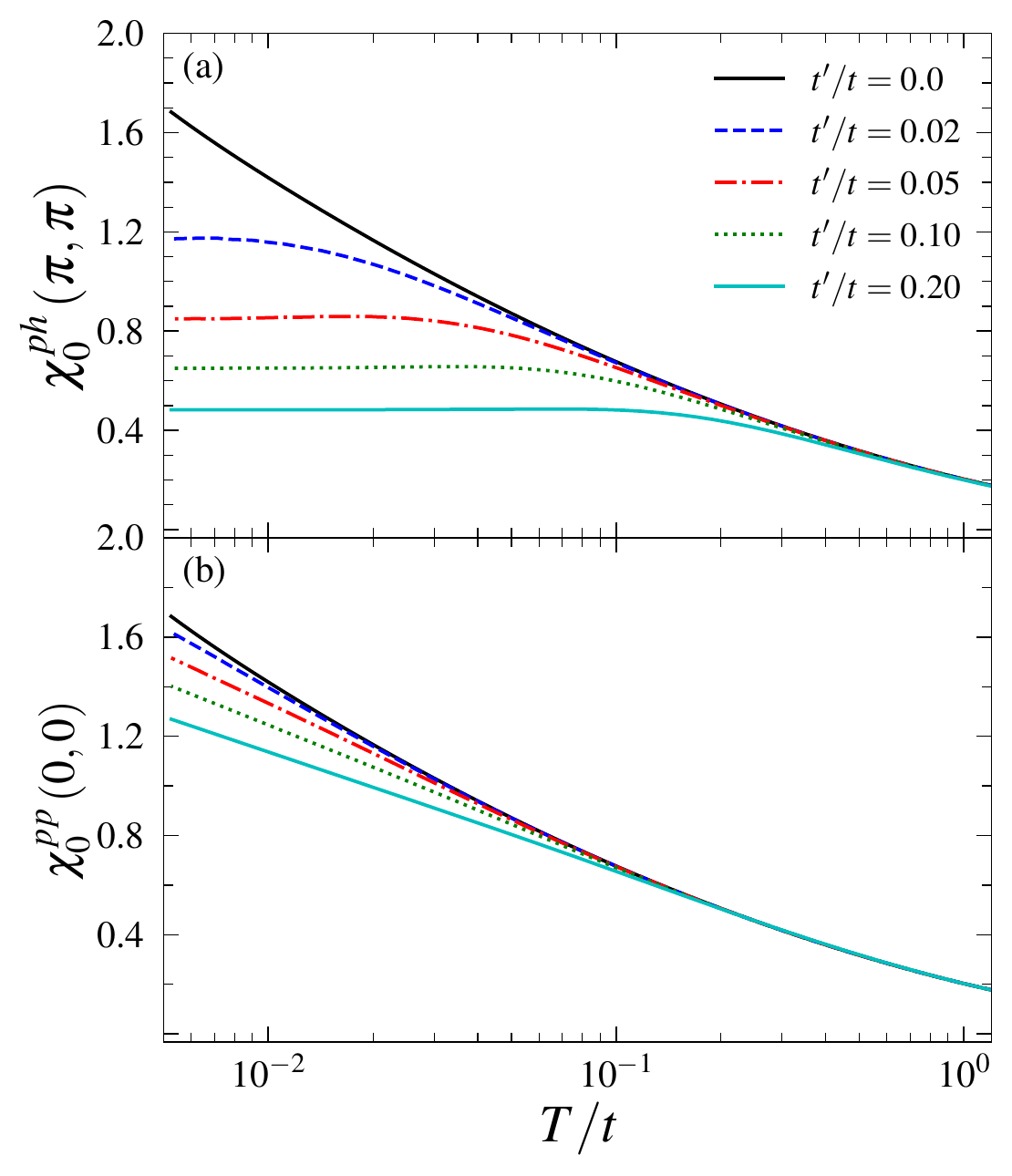}
\caption{Linear-log plots of the bare electronic susceptibilities for (a) the particle-hole and (b) the particle-particle channels as functions of temperature, and for several values of $t'$.}
\label{fig:chi0}
\end{figure}

Notice that the pseudo-spin operators are well-defined on any bipartite lattice. By fixing $\mu = 0$ and $t' = 0$, one can show that $\left[\mathcal{H},\eta^{z}\right] = \left[\mathcal{H},\eta^{\pm}\right] = 0$. Thus, as mentioned before, at half filling ($\mu=0$) and in the absence of NNN hopping, CDW and SS orders are exactly degenerate, forming an effective three-component pseudo-spin order parameter\,\cite{Yang90}. 
As a consequence of the Mermin-Wagner theorem, there is no long-range order at finite temperatures. By contrast, when $t' \neq 0$, one has $\left[\mathcal{H},\eta^{\pm}\right] \neq 0$, lifting the pseudo-spin SU(2) symmetry. In this case, one may expect the emergence of a finite temperature transition (either CDW or SS) even at half filling.

To further examine the impact of breaking the SU(2) symmetry when $t' \neq 0$, we analyze the behavior of the particle-hole ($\chi_{0}^{\rm ph}$) and particle-particle ($\chi_{0}^{\rm pp}$) bare electronic susceptibilities. For the noninteracting case, and focusing on the staggered particle-hole and the uniform particle-particle responses, they are defined as
\begin{align}
\chi_{0}^{\rm ph}(\mathbf{q}) &= -\sum_{\mathbf{k}} \frac{f(\varepsilon_{\mathbf{k}+\mathbf{q}}) - f(\varepsilon_{\mathbf{k}})}{\varepsilon_{\mathbf{k}+\mathbf{q}} - \varepsilon_{\mathbf{k}}}~,\\
\chi_{0}^{\rm pp}(0,0) &= \sum_{\mathbf{k}} \frac{1 - 2f(\varepsilon_{\mathbf{k}})}{2 \varepsilon_{\mathbf{k}}}~,
\end{align}
with $f(\varepsilon_{\mathbf{k}})$ being the Fermi-Dirac function, $\mathbf{q} = (\pi,\pi)$, and $\varepsilon_{\mathbf{k}}$ defined in Eq.\,\eqref{eq:ek}. 
Figures \ref{fig:chi0}\,(a) and \ref{fig:chi0}\,(b) show the temperature dependence of $\chi_{0}^{\rm ph}(\pi,\pi)$ and $\chi_{0}^{\rm pp}(0,0)$, respectively.
When $t' = 0$, both susceptibilities diverge logarithmically due to the van Hove singularity and perfect nesting. Once $t'$ is allowed, however, $\chi_{0}^{\rm ph}(\pi,\pi)$ is strongly suppressed and saturates to a finite value as the temperature decreases. The particle-particle susceptibility, on the other hand, continues to diverge logarithmically, displaying just a mild suppression relative to the $t' = 0$ case.

\begin{figure}[t]
  \centering
  \includegraphics[scale=0.53]{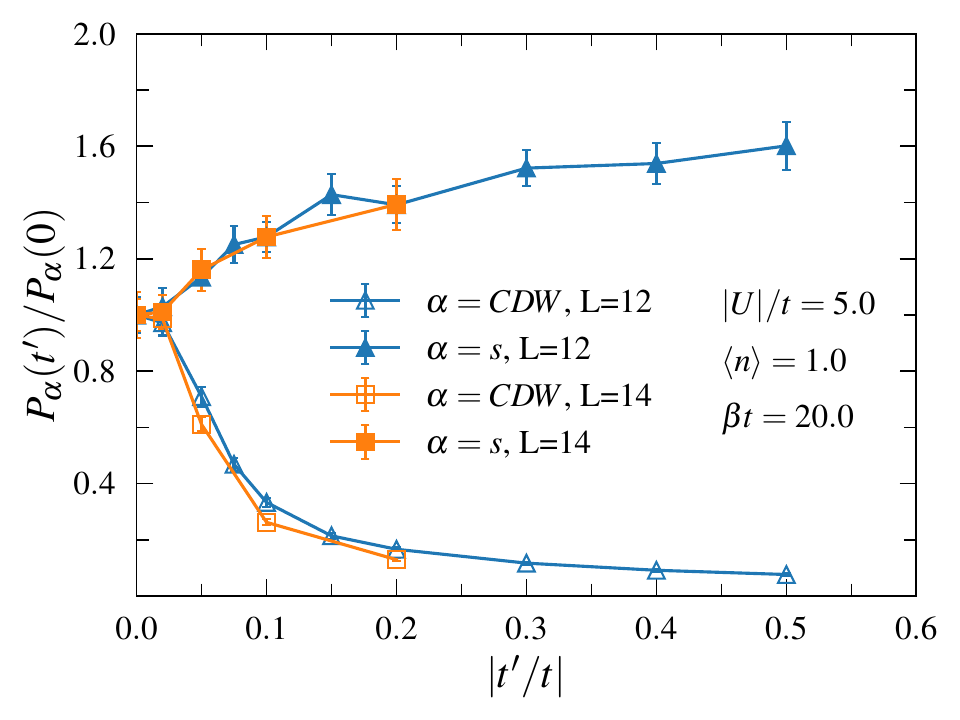}
   \caption{
   CDW (empty symbols) and $s$-wave pairing (s) (filled symbols) structure factors as functions of $|t'/t|$, normalized to their values at $t'=0$. 
   Data correspond to half filling, fixed $U=-5.0t$ and $\beta t = 20$, and linear lattice sizes $L=12$ (triangles)  and $14$ (squares).}
 \label{fig:Sa_CDW_SS_n10}
\end{figure}

Although the noninteracting behavior is relevant, it is also important to verify whether similar trends persist when $U \neq 0$, within QMC simulations. To this end, we examine the CDW and SS structure factors as functions of $|t'/t|$ at half filling, for $\beta t = 20$ and $U = -5t$, each normalized by its value at $t' = 0$. The resulting behaviors of $P_{\rm CDW}$ and $P_{s}$ are shown in Fig.\,\ref{fig:Sa_CDW_SS_n10} for system sizes $L = 12$ and $L = 14$. Notably, the data for $P_{\rm CDW}(t')$ show a strong suppression of CDW correlations as $t'$ increases, while $P_{s}(t')$ is enhanced. In other words, the inclusion of NNN hopping frustrates charge order, while enhancing SS correlations in the ground state -- in line with expectations from electron-phonon models \cite{Araujo2022}. Taken together, these results open the possibility of having a finite $T_c$ even at half filling, as discussed in the next subsection.

\begin{figure}[t]
  \centering
  \includegraphics[scale=0.48]{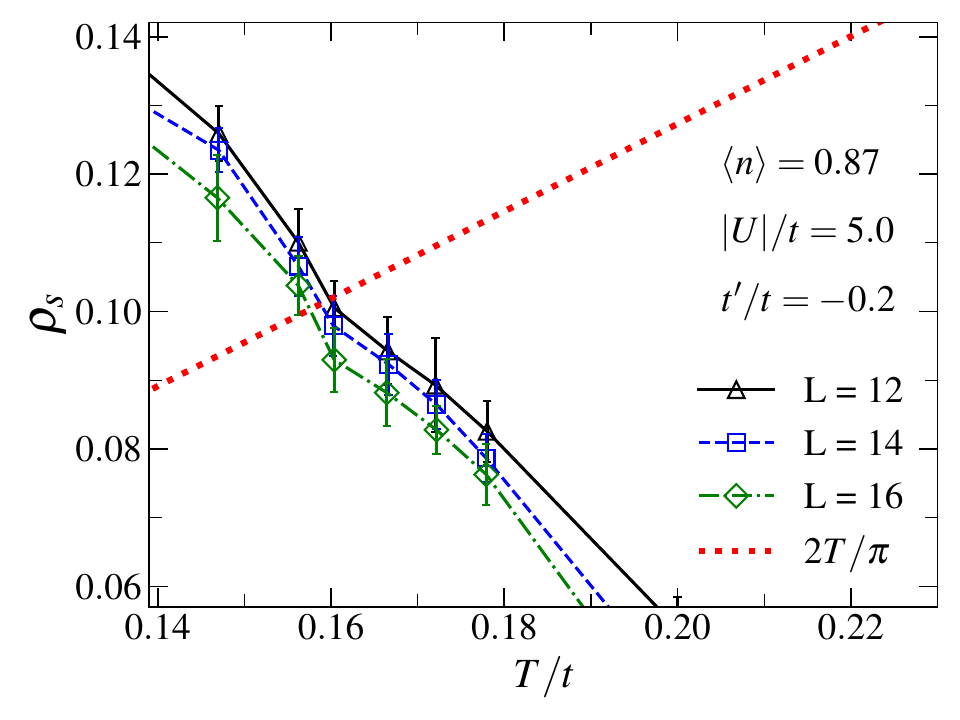}
   \caption{Temperature dependence of the superfluid density, $\rho_{s}$, for fixed $\langle n\rangle $, $t^{\prime}$ and $U$;  the curves are labelled by the lattice size, $L$.
   The (red) dotted line is $2T/\pi$.
}
 \label{fig:phos}
\end{figure}

\subsection{Critical Temperature}
\label{susec:Citical_Temperature}

For an accurate determination of $T_c$ we use the universal-jump relation for the helicity modulus, Eq.\,\eqref{eq:helicity_mod}.
Figure \ref{fig:phos} illustrates the procedure for $\langle n\rangle = 0.87$, for fixed $U = -5 t$ and $t^{\prime} = -0.2t$: For each linear size, $L$, we plot $\rho_s$ as a function of $T/t$, whose intersection with the straight line $2T/\pi$ yields an estimate for $T_c$. We note that the positions of the intersections are not too sensitive to $L$. 
Thus, the crossings with the straight line must occur within the same temperature range for all lattice sizes. Thus, either taking into account the scatter of all intersections, or just using the data for the largest lattice, we estimate $T_{c}/t = 0.164 \pm 0.002$.

\begin{figure}[t]
  \centering
  \includegraphics[scale=0.28]{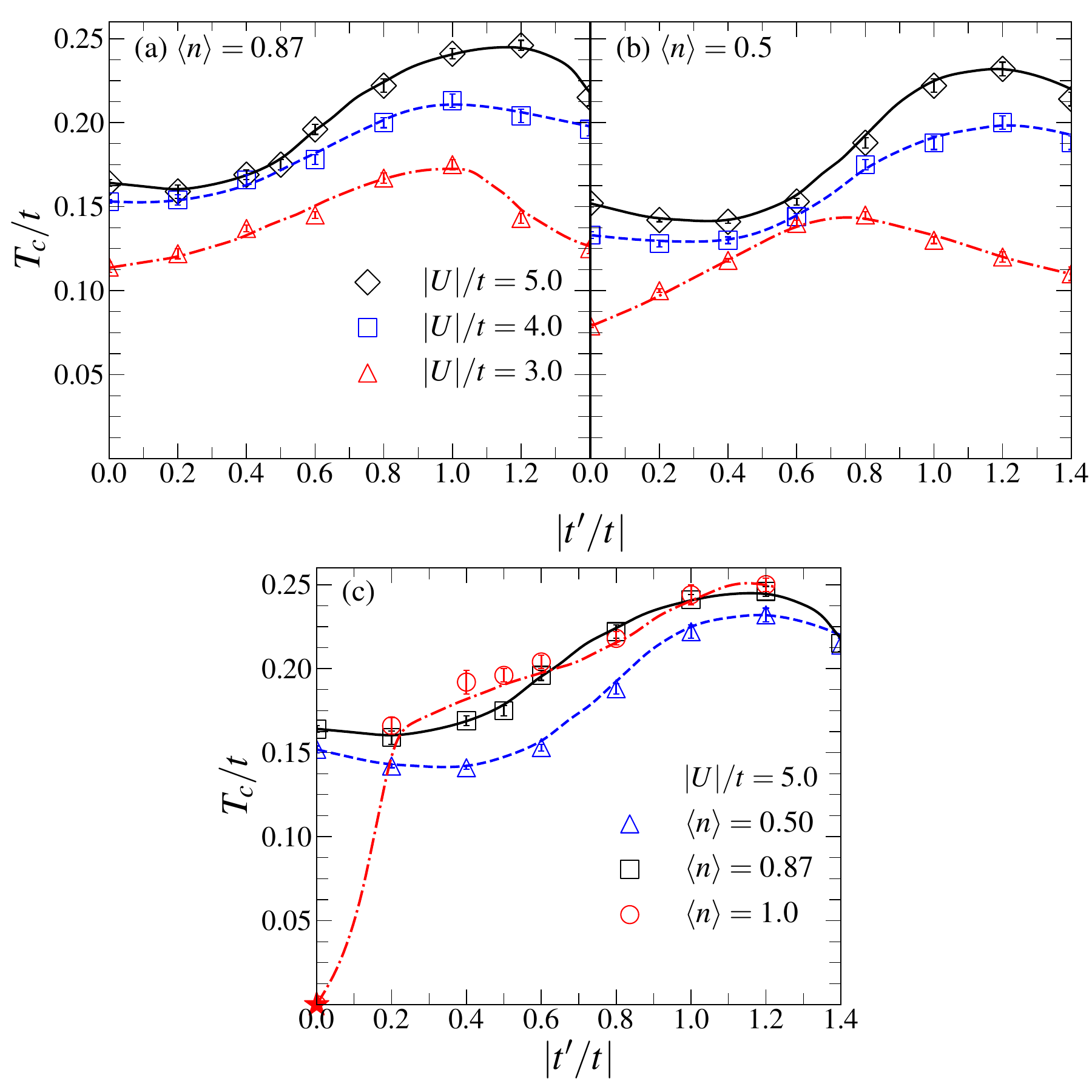}
   \caption{
   Critical temperature as a function of $|t'/t|$: 
   (a) for $\ave{n}=0.87$, and different values of $|U|/t$; 
   (b) same as (a), but for $\ave{n}=0.5$;
   (c) for $|U|/t=5$, and different values of $\ave{n}$.
   Lines through  points are guides to the eye, and most of the error bars are smaller than the data points.
   }
 \label{fig:Tc_tp_U_n087}
\end{figure}

\begin{figure*}[t]
  \centering
  \includegraphics[scale=0.39]{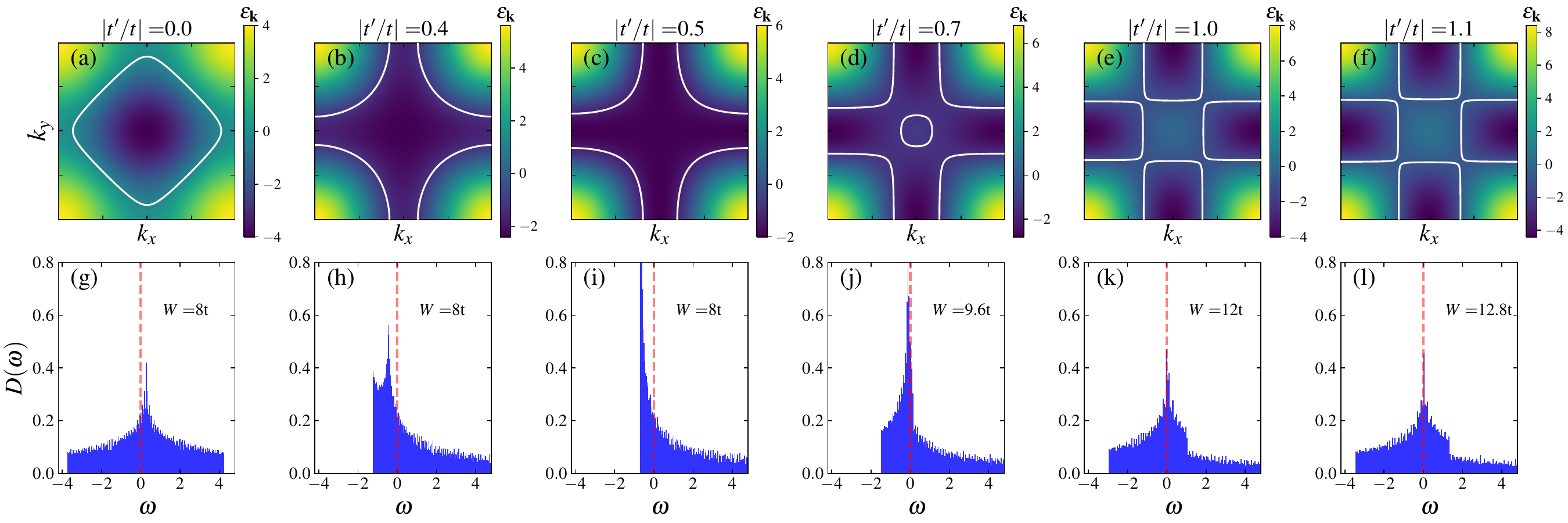}
   \caption{(a)-(f): Contour map of the single-particle dispersion, $\varepsilon_\kv$, for different values of $t'/t$; in each panel, the Fermi surface for $\ave{n}=0.87$ is shown as a white line. (g)-(l): corresponding evolution of the density of states; the dashed red vertical lines locate the corresponding Fermi energies. $W$ is the $t'$-dependent bandwidth, obtained from Eq.\,\eqref{eq:ek}.}
 \label{fig:DOS_FL}
\end{figure*}

We repeat this procedure for other values of $U$ and $t^{\prime}$, making use of the weak dependence of the superfluid density with $L$; accordingly, in what follows most of the results for $T_c$ have
been determined from simulations on lattices with linear size $L = 16$. 
Figure \ref{fig:Tc_tp_U_n087}\,(a) shows the behavior of the critical temperature as a function of $|t^{\prime}/t|$, for different values of $U$ and fixed $\langle n\rangle = 0.87$. 
For all values of interaction strength we analyzed, there is an increase in $T_c$ relative to its value for $t'=0$, $T_c^0$. 
Interestingly, the maximum $T_c$ occurs at slightly different values of $t^{\prime}$ when varying $|U|/t$. 
For the values of $U$ shown, the maximum critical temperature, $T_c(\langle n\rangle,|U|/t)/t = 0.246 \pm 0.003$, is reached for $|U|/t = 5.0$ and $t^{\prime} = -1.2t$. 
This represents a $50\%$ increase compared with the corresponding $T_c^0$. Similar results are obtained at quarter-filling, as shown in Fig.\,\ref{fig:Tc_tp_U_n087}\,(b), with a maximum $T_c(\langle n\rangle,|U|/t)/t = 0.232 \pm 0.003$ at $t'=-1.2t$; again an increase of about 50\% over the value for $t'=0$.
Overall, we see that $T_c$ increases with $|U|$, with $t'$ controlling how much the increase in $T_c$ is, relative to the value when $t'=0$.

At this point, it is instructive to examine the behavior of the non-interacting DOS and Fermi surface for different values of $t'$. 
For instance, Figs.\,\ref{fig:DOS_FL}\,(a)-(f) show the evolution of the Fermi surface with $t'$ for $\ave{n}=0.87$, 
where one may notice a change from electron-like  to hole-like behavior as $|t'|$ increases. 
The corresponding evolution of the density of states is shown in Figs.\,\ref{fig:DOS_FL}\,(g)-(l), from which one also notices that for $|t'/t|\lesssim0.5$, $D(0)$ is approximately constant, and in the interval $0.7 \lesssim |t'/t|\leq 1.1$ $D(0)$ lies at, or very close to, a van Hove singularity. 
We can therefore resort to Eq.\,\eqref{eq:Tc_BCS} to explain the slow increase of $T_c$ with $t'$ in the former interval, and a faster increase in the latter, as shown in Fig.\,\ref{fig:Tc_tp_U_n087}\,(a).
Also, this increase above $|t'/t| \approx 0.6$ is faster for larger $|U|$.
Another interesting aspect emerging from Fig.\,\ref{fig:DOS_FL} is that the bandwidth is constant up to $|t'| = 0.5t$, and increases with $t'$ thereafter, which indicates that care must be taken when normalizing energies by $W$.

We now turn to the behavior at half filling. As discussed earlier, any finite $t'$ breaks the pseudo-spin SU(2) symmetry by frustrating charge order and favoring pairing. Accordingly, we repeat the same analysis for the superfluid density at $\langle n \rangle = 1$ and compare it with doped cases. Figure~\ref{fig:Tc_tp_U_n087}\,(c) shows $T_c$ as a function of $|t'/t|$ for different band fillings $\langle n \rangle$, at fixed $U/t = -5$. Note that, except at $t' = 0$ -- where the Mermin-Wagner theorem enforces $T_c = 0$ at half filling -- the critical temperature at $\langle n \rangle = 1$ rises sharply with increasing $t'$, becoming the largest $T_c$ among all fillings considered.
Such a sharp increase occurs due to the strong suppression of the charge correlations for $t'\lesssim 0.1$, as displayed in Fig.\,\ref{fig:Sa_CDW_SS_n10}.
For large values of NNN hopping, both $\langle n \rangle = 1$ and $\langle n \rangle = 0.87$ display similar behavior, with a maximum around $t' = -1.2t$, which seems to be the global maximum critical temperature value in the system.

\subsection{Pairing Temperature}
\label{susec:Pairing_Temperature}

Unlike conventional superconductors, the high-$T_c$ cuprates display a pseudogap in the spectrum of spin excitations \cite{Bucher93,Valla06}, which is usually associated with the formation of Cooper pairs below some temperature scale $T_p \gtrsim T_c$; these preformed pairs condense into a superconducting state at $T_c$ \cite{Kordyuk15}. 
One of the signatures of this behavior is a drop in the uniform susceptibility, $\chi_s$, as the temperature decreases \cite{Randeria92,dosSantos94,Magierski09,Magierski11,Valla06,Paiva10, Kordyuk15, Wlazlowski13, Tajima14, Fontenele22}. Therefore, it is also instructive to determine the behavior of $T_p$ as a function of $t'$ in our model.

\begin{figure}[t]
  \centering
  \includegraphics[scale=0.36]{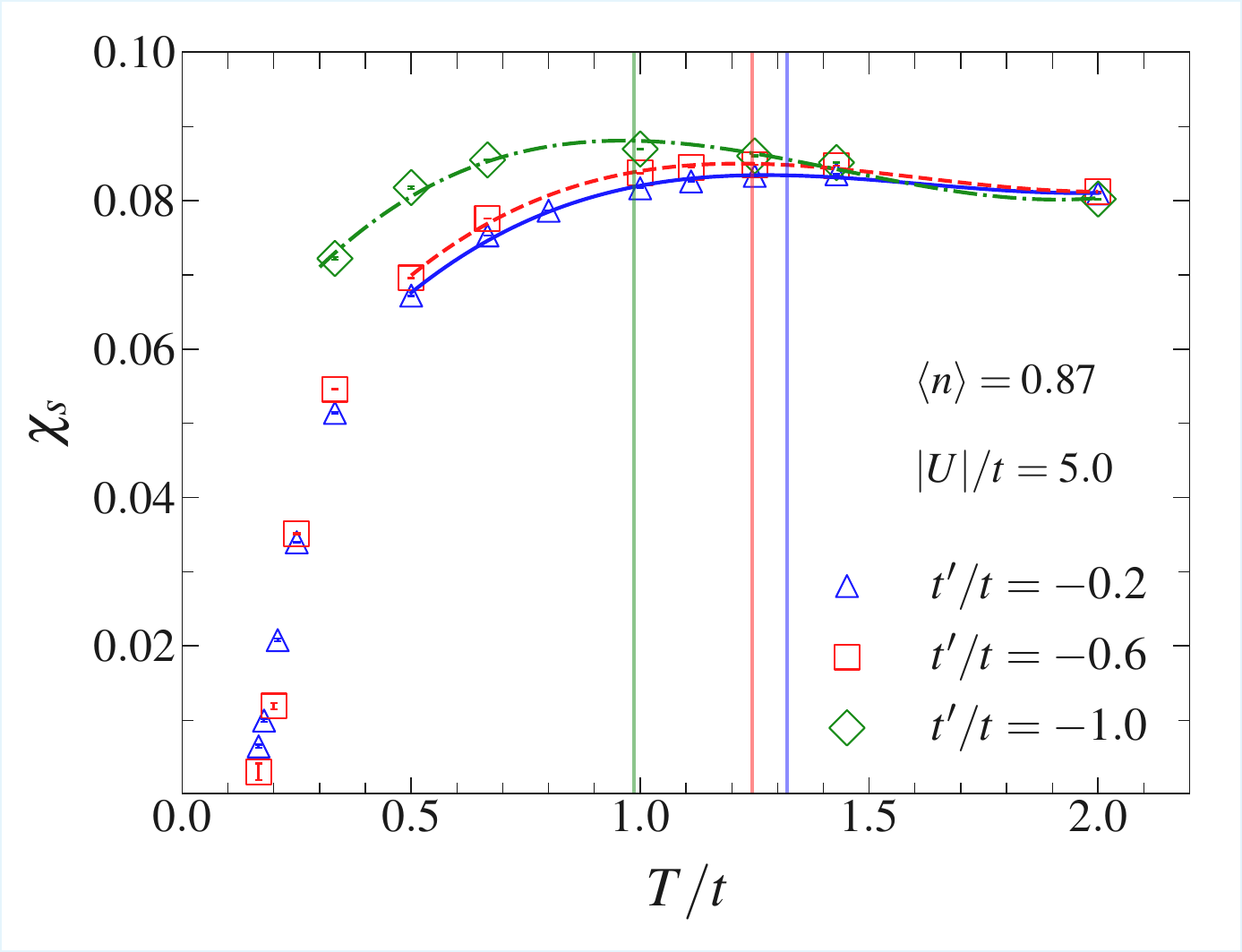}
   \caption{Uniform spin susceptibility as a function of temperature, for different values of $t^{\prime}$, with fixed $\ave{n} = 0.87$ and $|U|/t=5.0$, for a linear lattice size $L=16$. 
   The curves through data points represent polynomial interpolations, and the vertical lines locate the maxima.  
   The error bars are smaller than data points. }
\label{fig:Tp_n05U3}
\end{figure}

\begin{figure}[t]
  \centering
  \includegraphics[scale=0.45]{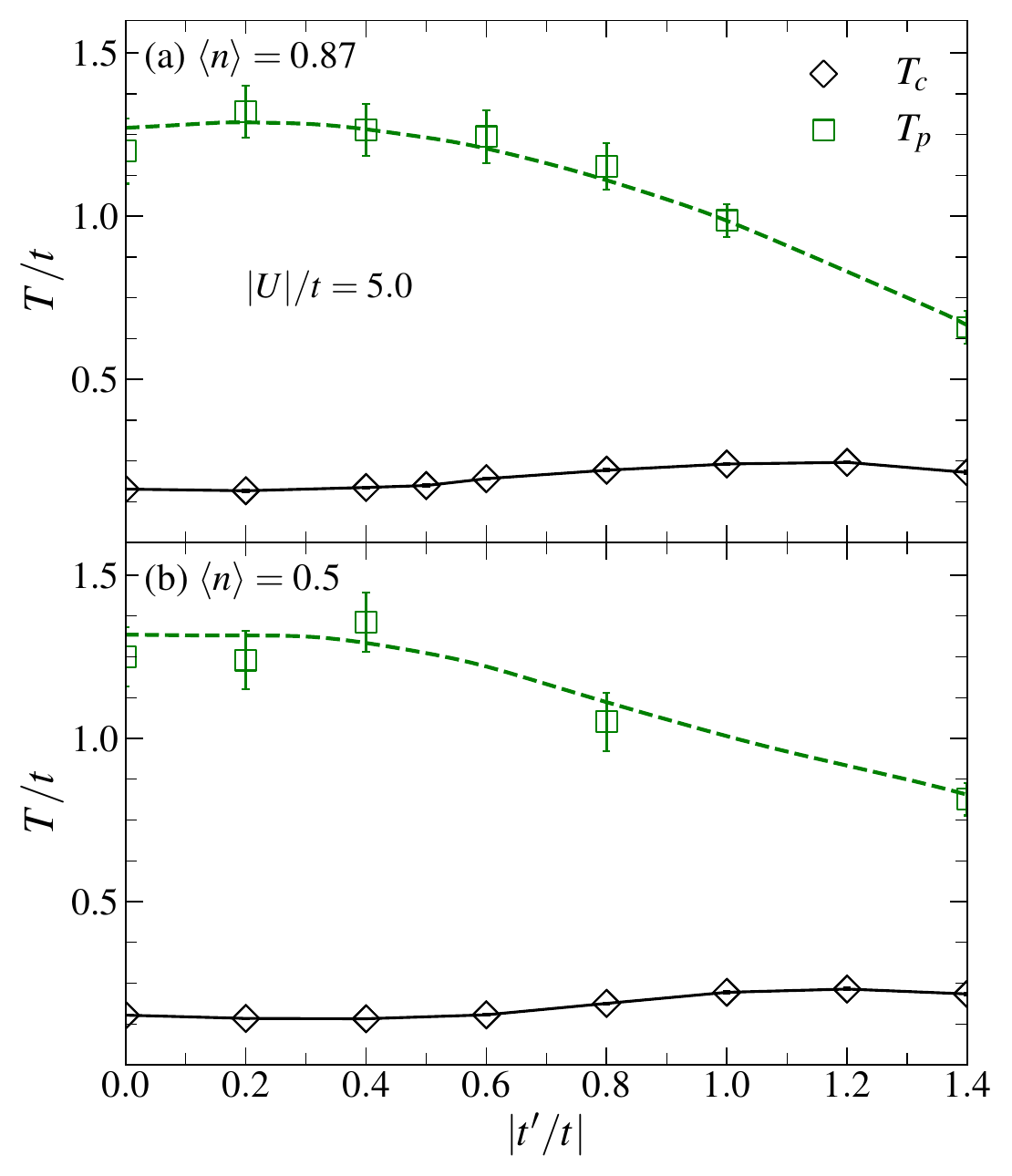}
   \caption{Pairing temperature $T_{p}$ (squares), and critical temperature $T_{c}$ (diamonds), as functions of $t'$, for fixed   $|U|/t=5.0$: (a) $\ave{n} = 0.87$ and (b) $\ave{n}=0.5$. 
   Lines are guides to the eyes. }
 \label{fig:Tc_Tp_n05n087_U50}
\end{figure}


\begin{figure}[t]
  \centering
  \includegraphics[scale=0.34]{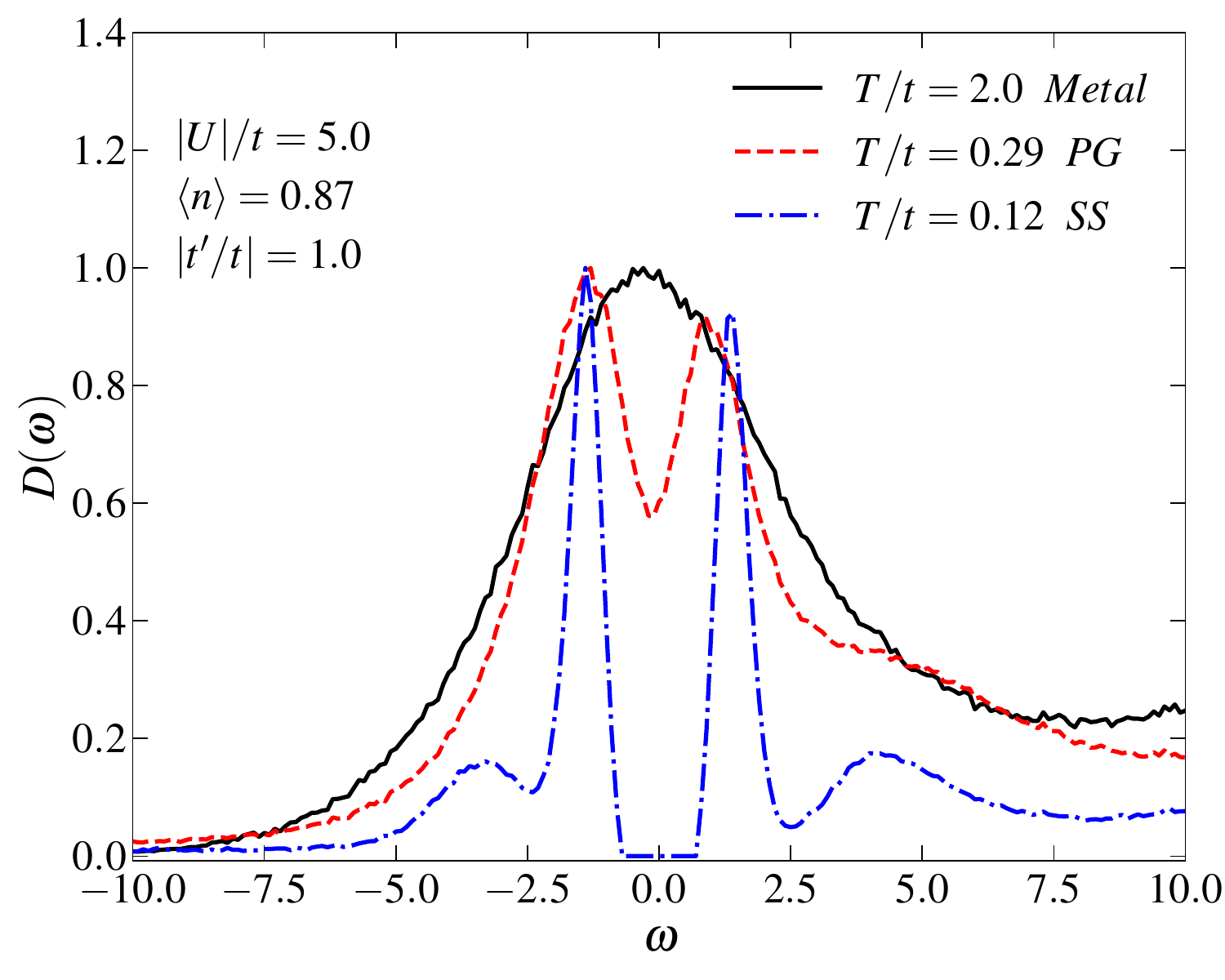}
   \caption{Interacting DOS at different temperatures for fixed $\ave{n}=0.87$, $|U|/t = 5.0$ and $|t^{\prime}/t|=1$, for a $16\times 16$ lattice. 
   PG and SC respectively stand for pseudogap and superconducting phases.} 
  \label{fig:DOS_n087_U50_tp1_Tdif}
\end{figure}

\begin{figure}[t]
  \centering
  \includegraphics[scale=0.2]{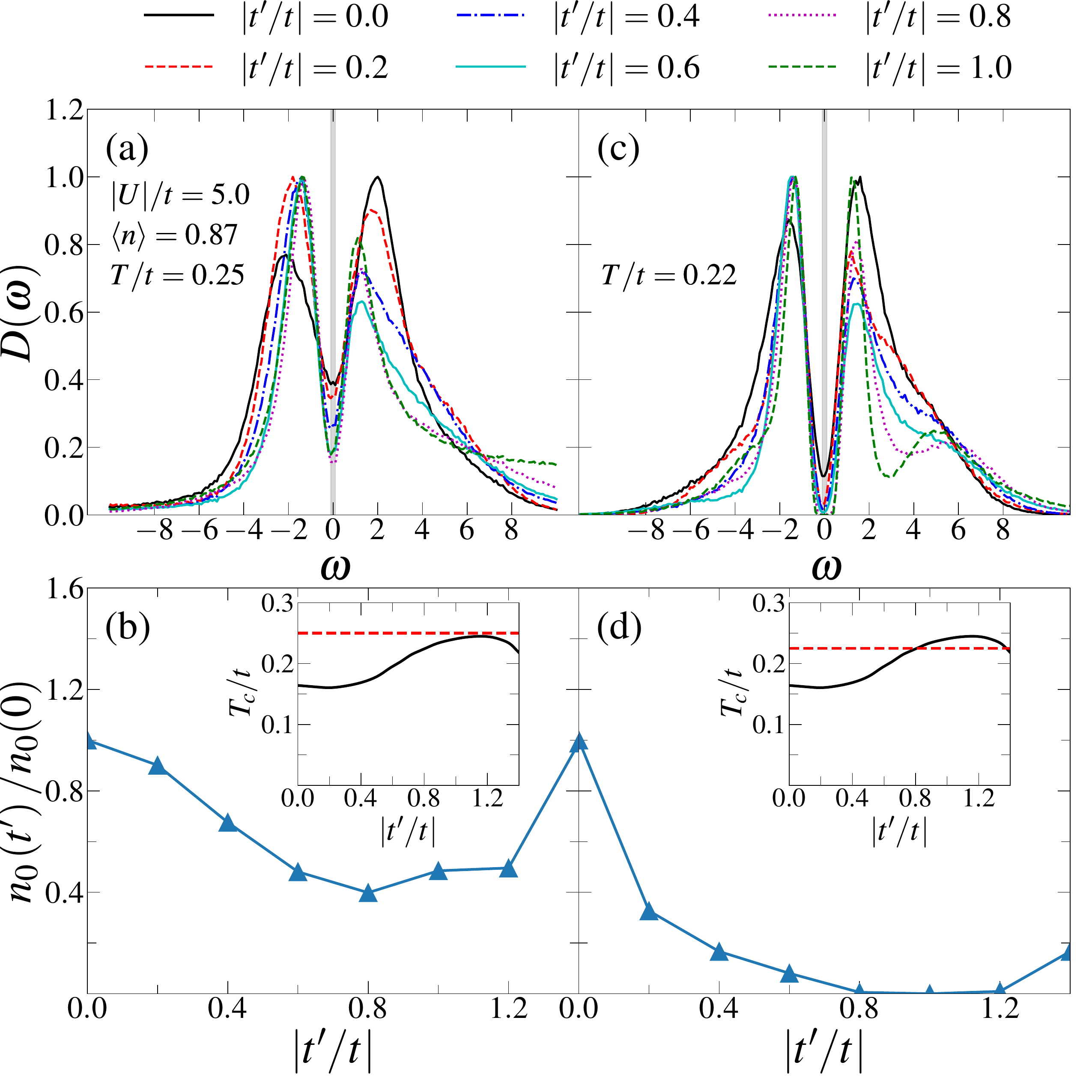}
    \caption{(a) Interacting density of states, $D(\omega)$, for different values of $t'$, and fixed $U/t = -5.0$, $\ave{n} = 0.87$. The temperature is kept fixed at $T/t=0.25$.
   (b) Fermionic density, $n_0$ [Eq.\,\eqref{eq:nfl_w}], within an energy range $2\Delta$ around $\omega=0$. We take $2\Delta=0.2t$, as represented by the shaded area around $\omega=0$ in panel (a). The inset shows that $T_c$ lies below the temperature used to calculate $D(\omega)$ in panel (a), shown as a (red) dashed horizontal line.
   (c) Same as (a), but for $T/t=0.22$.
   (d) Same as (b) but for $T/t=0.22$, which now crosses $T_c$.} 
 \label{fig:nFL_n087}
\end{figure}


Figure \ref{fig:Tp_n05U3} exhibits DQMC data for the temperature dependence of the uniform susceptibility, $\chi_s$ [see Eq.\,\eqref{eq:magsusc}], for different values of $t^\prime$, and for fixed $|U|/t = 5.0$ and $\langle n \rangle = 0.87$. 
For each $t^\prime$ we see that $\chi_{s}$ drops steadily below some temperature. 
In line with the idea that this downturn signals the formation of local pairs within some temperature scale~\cite{Randeria92,dosSantos94,Magierski09,Magierski11,Wlazlowski13,Tajima14}, we adopt the position of the maximum in $\chi_s$ as the pairing scale, $T_p(t^{\prime})$. 
As Fig.\,\ref{fig:Tp_n05U3} shows, the maxima can be quite broad, so that their positions are determined by performing a polynomial fit through the data.
The error bars are estimated as half of the interval between two temperatures around the maxima. 
This somewhat flexible definition is consistent with the fact that we are dealing with a crossover, hence $T_p$ must be regarded as signaling a temperature scale, not a sharp transition.

We repeat this procedure for other values of $t'$, and collect the estimates for $T_p(t')$ in Fig.\,\ref{fig:Tc_Tp_n05n087_U50}\,(a), which also includes the corresponding data for $T_c$ from Fig.\,\ref{fig:Tc_tp_U_n087}\,(a), for comparison. 
Figure \ref{fig:Tc_Tp_n05n087_U50}\,(b) shows similar plots for $\ave{n}=0.5$. 
It is interesting to note that increasing $t'$ causes a decrease in $T_p$, and an increase in $T_c$: the NNN hopping channel tends to suppress pair formation, while favoring their condensation. 
The merging of $T_p$ and $T_c$ for large $|t'/t|$ also suggests a route to approach the BCS limit even at moderate coupling, which, when $t'=0$, occurs at $|U/t|\ll1$.   

Further insight into the behavior within this metal-superconductor transition, crossing a preformed-pair region in the process, may be obtained through the interacting DOS, $D(\omega)$. 
The latter is obtained from data for the imaginary-time Green's function, $G(|\mathbf{i}-\mathbf{j}|,\tau)$, extracted from our DQMC simulations, and using the maximum-entropy method \cite{Jarrell96} to invert the integral equation,
\begin{align}
       G(0,\tau) &= \int d\omega D(\omega) \frac{e^{-\omega\tau}}{1 + e^{-\beta\omega}}.
       \label{eq:DOS_int}
\end{align}

Figure \ref{fig:DOS_n087_U50_tp1_Tdif} shows $D(\omega)$ for $\ave{n} = 0.87$, with fixed $U/t =~-5$, $|t'/t| = 1.0$, and linear size $L=16$. For $T/t=2$, the system is in a metallic gapless phase, and $D(0)$ is maximum. As the temperature is decreased, states tend to be suppressed at the Fermi energy, $\omega=0$, which is associated with the pseudogap region. Further decrease in the temperature leads to a complete depletion of states at $\omega=0$, when the gap opens and the system becomes superconducting. This scenario is in line with previous findings for the AHM with $t'=0$ \cite{Singer99}.

Let us now investigate what happens to $D(\omega)$ around the Fermi energy, $\varepsilon_\text{F}$, at temperatures near $T_c$. 
To this end, we examine the electronic density within a very narrow range, $2\Delta$ around $\omega=0$, defined as 
\begin{align}
    n_{0} = \lim_{\Delta \rightarrow 0} \int_{-\Delta}^{\Delta} d\omega D(\omega).
    \label{eq:nfl_w}
\end{align}
This quantity must vanish when the system enters the SC phase, due to the opening of the gap at $\varepsilon_\text{F}$. 
In Fig.\,\ref{fig:nFL_n087}\,(a) we show $D(\omega)$ for several values of $t^{\prime}$, with $|U|/t$ and $\ave{n}$ kept fixed; the temperature $T/t$ is constant, chosen to be slightly above the maximum $T_c$ for all values of $t'$ considered [see the inset in Fig.\,\ref{fig:nFL_n087}\,(b)].  
We see that $D(0)$ never vanishes. 
Most interestingly, Fig.\,\ref{fig:nFL_n087}\,(b) shows that $n_0$ first decreases steadily with $|t'/t|$, signaling a continuing depletion of the Fermi level. However, this depletion is incomplete, since the temperature is always higher than $T_c$, so that $n_0$ never vanishes. 
Beyond the maximum $T_c$, $n_0$ starts increasing, indicating a rise in the occupation of the Fermi level. 
One may therefore conclude that $n_0$ at a given temperature measures how far one stands from $T_c$, as the remaining parameters are varied.

By contrast, Figs.\,\ref{fig:nFL_n087}(c) and \ref{fig:nFL_n087}(d) show the corresponding data at a temperature slightly lower than the maximum $T_c$. In this case, we see that $D(0)$ actually vanishes, with a finite superconducting gap opening. Accordingly, $n_0$ now reaches zero, indicating a complete depletion of the Fermi level. This persists until $|t'/t|$ is such that $T\gtrsim T_c$, allowing an increasing occupation of the Fermi level. This reinforces the role of $n_0$ as a measure of distance from $T_c$ within the pseudogap region.

\section{Conclusions}
\label{sec:conc}

We have investigated how the presence of NNN hoppings affects pairing properties of the attractive Hubbard model on a square lattice. 
Through quantum Monte Carlo simulations, we have found that an increase in $|t'/t|$ can enhance $T_c$ up to 50\% for $|U|/t \lesssim 5$. 
However, the increase in $T_c$ is not monotonic, since a maximum value is achieved at some ratio $|t'/t|\sim 1$ which depends on both $U$ and the density $\ave{n}$.
Interestingly, since the inclusion of $t'$ breaks the pseudo-spin SU(2) symmetry by frustrating CDW order and favoring the SS phase, a finite-temperature phase transition emerges even at half-filling, exhibiting the largest $T_c$ compared to other fillings. The enhancement for $T_c(t')$ has its maximum at $|t'/t| \approx 1.2$, at fixed $U = -5t$ and $\langle n \rangle = 1$ or $0.87$, reaching $T_c \approx 0.25t$.

We have also examined the influence of $t'$ on the pairing temperature, $T_p$, which defines the region of pre-formed pairs (often associated with a pseudo-gap phase) above the superconducting phase. 
It turned out that $T_p$ behaves oppositely to $T_c$: it decreases with increasing $|t'/t|$, thus decreasing the temperature interval in which pre-formed local pairs exist without coherence. 
One may conjecture that for sufficiently large $|t'/t|$ the two temperatures merge, thus restoring the BCS regime even in the presence of a sizeable on-site coupling. 
We attribute this to the fact that while next-nearest neighbor hopping opens new paths for pairs to move, it also hinders pair formation since unpaired fermions also have more paths to follow.

We have also provided numerical evidence that the pseudogap indeed turns into a superconducting gap, by examining the evolution of the interacting density of states, $D(\omega)$. 
At two fixed temperatures, one slightly above and the other slightly below the maximum $T_c$, we clearly identify the dip in $D(0)$ vanishing within a finite energy interval when $T<T_c$.
In this context, we have introduced $n_0$, the density of fermions within a small interval, $\Delta$, around the Fermi energy. 
It turned out that $n_0$ at a fixed temperature provides  a measure of the distance from $T_c$ as the remaining parameters are varied.

In closing, we hope our results will stimulate the OLE's community to pursue further experiments to probe these effects of further neighbor hopping.

\section*{ACKNOWLEDGMENTS}
The authors are grateful to the Brazilian Agencies Conselho Nacional de Desenvolvimento Cient\'\i fico e Tecnol\'ogico (CNPq), Coordena\c c\~ao de Aperfei\c coamento de Pessoal de Ensino Superior (CAPES), and Instituto Nacional de Ci\^encia e Tecnologia de Informa\c c\~ao Qu\^antica (INCT-IQ) for funding this project. 
We also gratefully acknowledge support from Funda\c c\~ao Carlos Chagas de Apoio \`a Pesquisa (FAPERJ), through the grants E-26/200.258/2023 (N.C.C.),  
E-26/200.959/2022  (T.P.), 
E-26/210.100/2023  (T.P.), 
and E-26/204.333/2024 (R.R.d.S.).
We also gratefully acknowledge support from CNPq through the grants 
313065/2021-7 (N.C.C.),
308335/2019-8 (T.P.), 
403130/2021-2 (T.P.), 
442072/2023-6 (T.P.), 
and 314611/2023-1 (R.R.d.S.).



\bibliography{ref}
\end{document}